\documentstyle{elsart}
\begin{document}

\begin{frontmatter}
\title{Properties of Valence Nucleon Distributions for Halo Nuclei}
\author{C.J. Lin, H.Q. Zhang, Z.H. Liu, Y.W. Wu, F. Yang, and M. Ruan}
\address{China Institute of Atomic Energy, P.O. Box 275(10), Beijing, 102413, China}
\begin{abstract}
\quad With the binding energies and configurations determined experimentally, 
the root-mean-square radii are calculated for a number of single-particle states 
by numerically solving the Sch{\"o}rdinger equations. By studying the relations 
between the radii and separation energies, the new scaling laws and necessary 
conditions for neutron halos and proton halos are established, respectively. 
Especially the existence region of true proton halos is pointed out. It is found 
that the effects of short-distance behaviours of valence nucleons at the edges 
of interaction potentials can not be disregarded. Moreover, by means of the radii 
of interaction potentials, the contributions of outer parts are estimated as 
the criterions of halos.
\end{abstract}
\begin{keyword}
Scaling laws for valence neutron and proton distributions in halo nuclei; 
Single-particle potential model; 
Criterion for halo occurrence.
\PACS 21.10.-k; 21.60.-n; 27.20.+n; 27.30.+t
\end{keyword}
\end{frontmatter}

\quad The exotic properties of halo nuclei have been extensively investigated 
since this phenomena were found by Tanihata et al.~\cite{tani01,tani02}. 
So far, a number of light, unstable nuclei are experimentally found 
to be halo states, i.e., the one-neutron halo nuclei, $^{11}$Be~\cite{fuku01} 
and $^{19}$C~\cite{bazi01}; the two-neutron halo nuclei, $^6$He, 
$^{11}$Li~\cite{tani01,tani02,tani03}, $^{14}$Be~\cite{suzu01,labi01}, and 
$^{17}$B~\cite{suzu01}; the four neutron structure, 
$^8$He~\cite{tani01,tani02,zhuk01}; the proton halo nuclei, 
$^8$B~\cite{smed01,fuku02,guim01}, $^{17}$F, $^{17}$Ne~\cite{ozaw01}, 
$^{20}$Mg~\cite{suzu02}, and $^{26,27,28}$P~\cite{navi01}; and the one-neutron 
halo structures in excited states of $^{12}$B and $^{13}$C~\cite{liu01,lcj01}; 
and so on. Meanwhile, the general properties of halo nuclei have been gradually 
understood~\cite{hans01,hans02,fedo01,jens01}. Nowadays, ones have recognized 
that in principle halo is a threshold effect when the state close to the 
continuum~\cite{hans02}. The necessary conditions to form halos are weak
bound, low angular momentum, and short-range interaction. And these conditions 
are formulated as general scaling rules~\cite{fedo01,jens01}. The general 
properties of halos in the above works, for example the scaling laws, are 
extracted by means of the square well potentials, because the Sch{\"o}rdinger 
equations have the analytic solutions. Fedorov et al.~\cite{fedo01} found that
some results of Gaussian and surface potentials obey the scaling laws of square 
potentials. Moreover, Hansen et al.~\cite{hans02} 
drew two qualitative conclusions by studying the asymptotic behaviours of radial 
wavefunctions for two-body systems, i.e. (1) the states with $l=2$ can not form 
the halo structures due to the affections of the centrifugal barriers; (2) the 
proton halos do not exist due to the affections of the long-range Coulomb potentials. 
Nevertheless, we often find that the experimental findings are not in agreement 
with these scaling laws. Of course, the asymptotic behaviours are the same for 
different potentials. But the short-distance behaviors of the valence nucleons 
at the edges of the interaction potentials should affect these conclusions. In 
other words, the results depend on the potential shapes. In this letter, we 
re-focus on the scaling laws of valence nucleon distributions: According to the 
experimental findings, adopt the more real shapes, such as Woods-Saxon forms, 
to get the more realistic scaling laws, and to reveal the affections of the 
short-distance behaviours.

\quad For the sake of simplicity, the two-body systems are considered and the 
single-particle potential model\cite{fedo01,sher01} is employed presently. 
In principle, it supposes that the nucleus is configurated by a core with a 
valence nucleon outside. In present work, the configuration of certain 
single-particle bound state is assigned by the experimental results, for example, 
the binding energy, angular momentum, spin, etc.. The Sch{\"o}rdinger equation 
of this state can be solved by the numerically analytic method. Then the radial 
wavefunction, the density distribution, and root-mean-square (rms) radius etc., 
can be calculated. The following points need to be mentioned: 
(1) Adopt the more real shape for the nuclear potential, i.e. Woods-Saxon shape,  
$f(x)=[1+exp(x/a)]^{-1}$ with $x=r-r_{0}A_{c}^{1/3}$, $r_{0} = 1.25$ fm, and 
$a = 0.65$ fm; (2) For the cases of which valence nucleons are protons, 
the Coulomb potential of a uniformly charged sphere with radius 
$r_{0C}A_{c}^{1/3}$ and $r_{0C}=1.30$ fm is added in the interaction 
potential; (3) The spin-orbit-coupling potential with Thomas form 
$\lambda=25$~\cite{satc01} is also considered, because it is influential for 
some states with high angular momenta ($l>2$)and spins. Furthermore, the nuclear 
potential depth $V$ is determined by the binding energy and the 
spin-orbit-coupling potential, $V_{s.o.}$ is varied with $V$, i.e. 
$V_{s.o.}=(\lambda / 45.2)V$. All the parameters mentioned above are commonly 
used as 'standard parameters'. Sherr~\cite{sher01} pointed out that the predicted 
radii are insensitive to moderate changes of these parameters. It ascribes to 
the gross results of integrations of radial wavefunctions. Whereas the wavefunctions, 
describing the detail behaviours of the valence nucleons, are sensitive to some 
extent. Fig.~\ref{fig1} illustrates the single-particle wavefunctions for 
two bound states $2s_{1/2}$ and $1p_{1/2}$ of $^{11}$Be with Woods-Saxon form 
factors with the standard parameters (solid lines) and with the other parameters 
($r_{0}=1.05$ fm, $a=0.45$ fm; dashed lines), respectively. The results of the 
square well with $r_{0}=1.25$ fm (dotted lines) are also shown. The dash-dotted 
line indicates the position of the radius of interaction potential between core 
and valence nucleon. 

\quad  For the neutron cases, the current scaling laws~\cite{hans02,fedo01} 
of the square well are,
\begin{equation}
    \frac{<r_n^2>}{R_{cn}^2} \simeq \left\{\begin{array}{ll}
    			10.44 MeVfm^2/(S_nR_{cn}^2) & \ \ \ \ l=0 \\
    			3.65 MeV^{1/2}fm/(S_nR_{cn}^2)^{1/2} & \ \ \ \ l=1 \\
    			1.40 & \ \ \ \ l=2
                  \end{array} \right. ,
\end{equation}
where $R_{cn}$ is the radius of interaction potential (core + nucleon), $<r_n^2>$ 
and $S_n$ are the rms radius and separation energy of the valence neutron, 
respectively. According to the halo definition, i.e. the probability of finding 
the valence nucleon is larger than 50\% outside the range of the interaction 
potential~\cite{hans02,fedo01,jens01}, these states with $<r^2>/R_{cn}^2 > 2$ are 
the true halo states in strict meaning. The existence condition of neutron halo 
is derived form the above equation as yet. Obviously, no halo can be formed in 
the $l=2$ state.

\quad Using the single-particle potential model mentioned above, we calculated the 
rms radii of the valence neutrons for the neutron-rich nuclei from $^4$He to 
$^{22}$C, including some excited states. All the states calculated are dominated 
by single-particle states and their configurations are determined by the experiments. 
As the results, the solid symbols shown in the Fig.~\ref{fig2} illustrate the 
$<r_n^2>/R_{cn}^2$ versus $S_nR_{cn}^2$ for these states. Where the interaction 
potential radius is taken as, $R_{cn}=1.25(A_{core}^{1/3}+A_{val}^{1/3})$, with 
$A_{core}$ and $A_{val}$ are the mass numbers of the core and valence nucleon, 
respectively. It is worthy of being pointed out that such $R_{cn}$ definition 
is consistent in value with its in the Eq. 1, i.e. 
$R_{cn}^2=5/3(<r^2>_{core}+4fm^2)$. For example, for the $^{11}$Be nucleus, 
$<r^2>^{1/2}_{core} = 2.37$ fm~\cite{sher01} then $R_{cn} = 4.00$ fm, 
and $R_{cn} = 3.95$ fm in our case.
In the Fig.~\ref{fig2}, the solid lines are the fit results for the different 
orbit angular momenta. The forms of the fits are $y=a/x^b$, where $a$ and $b$ are 
the parameters. So the scaling laws given by the Woods-Saxon shapes are,
\begin{equation}
    \frac{<r_n^2>}{R_{cn}^2} \simeq \left\{\begin{array}{ll}
    			13.5 / (S_nR_{cn}^2)^{0.667} & \ \ \ \ l=0 \\
    			4.38 / (S_nR_{cn}^2)^{0.446} & \ \ \ \ l=1 \\
    			1.43 / (S_nR_{cn}^2)^{0.180} & \ \ \ \ l=2
                  \end{array} \right. .
\end{equation}
Compared with Eq. 1, it is immediately clear that they are quite different. In 
order to show the discrepancy, the scaling laws given by the square potentials 
are plotted as the dashed lines in the Fig.~\ref{fig2}. Then the necessary 
conditions for the formations of neutron halos are,
\begin{equation}
    S_nR_{cn}^2 < \left\{\begin{array}{ll}
    			17.6MeVfm^2 & \ \ \ \ l=0 \\
    			5.79MeVfm^2 & \ \ \ \ l=1 \\
    			0.153MeVfm^2 & \ \ \ \ l=2
                  \end{array} \right. .
\end{equation}
The actual conditions are more looser than those given by the square potentials. 
For example, there exist the possibilities of halo occurrences in the $l=2$ 
states, although the conditions are too rigour. Without doubts, the conditions 
expressed by Eq. 3 have the realistic meanings for searching the unknown neutron 
halo states.

\quad By introducing the Coulomb potentials, the density distributions of valence 
protons are calculated for some typical proton-rich nuclei, for example $^8$B, 
$^{17}$F, and some proton-rich isotopes of Z=15, 16, etc., also including some 
excitation states with configuration ascertained. As same as the Fig.~\ref{fig2}, 
the Fig.~\ref{fig3} shows the relations between the rms radii of valence protons, 
$<r_p^2>$ and the separation energies, $S_p$. The solid symbols are calculation 
results. The solid lines are the fit results, 
\begin{equation}
    \frac{<r_p^2>}{R_{cn}^2} \simeq 1.73/(S_pR_{cn}^2)^{0.246} \ \ \ \ l=0, 1.
\end{equation}
For the $l=2$ states, the data points are too few to get a assurable fit, i.e. the 
dotted lines shown in the Fig.~\ref{fig3}. So we do not discuss it here. It is found  
that the scaling laws are the same for the $l=0$ and $l=1$ states. We think it is 
just an accident, because of the affections by the long-range Coulomb potentials at 
the edges of the nuclei. Directly judging form Eq. 4, The necessary condition for 
the formation of proton halo is $S_pR_{cn}^2 < 0.551$ MeV. It should be noted that 
due to the existence of Coulomb barrier, some resonance states of which the energies 
are positive but lower than the barrier heights have enough surviving time and are 
able to form the halo structures. In other words, the proton halos could exist in 
these states. Hence, the actual necessary condition for the formation of proton 
halo is,
\begin{equation}
    -0.551/R_{cn}^2 MeV < E_p < V_B \ ,
\end{equation}
where $E_p$ is the proton energy, $V_B$ the height of Coulomb barrier. The shadow area 
in the Fig.~\ref{fig4} shows the region of proton halos occurrence. According to this 
condition, the existing findings figured as proton halos are in fact the proton skins. 
The location of $^{17}$F(2s$_{1/2}$) state, the most close to the proton halos region, 
is also illustrated in the figure. The actual conditions and the occurrence region 
should be useful to find out the true proton halos.

\quad Although the halo phenomenon is well studied, it is still confusion to judge that 
a nuclear state is a halo state or not experimentally. In general, when the distribution 
of a valence particle has a long tail, ones say that this nucleus has the halo structure. 
Such judgment is too ambiguous. Someone may think it is no meaning to make a strict 
definition for halo. But recent investigation indicates that because of the occurrence 
of neutron halo, there universally exists the overturn of the sequence of energy levels 
predicted by the shell model. It is well known that the shell model with harmonic-oscillator 
potential predicts that the $2s_{1/2}$ energy level is higher than the $1d_{5/2}$. 
But due to the halo state occurs in the $2s_{1/2}$ state, in fact its energy is lower 
than the $1d_{5/2}$. As an example, the first excited state of $^{13}$C is $2s_{1/2}$ 
state, not $1d_{5/2}$. Such overturn phenomenon is the signature of halo occurrence 
and results in the emergence of new magic number, N=16~\cite{ozaw02}. So it is necessary 
to define a halo state. Clearly, the nucleon distributions outside the nuclear 
cores are important for halo structures. The contributions of the outer part can be 
estimated by~\cite{liu01,lcj01,cars01}
\begin{equation}
    D_{\lambda}=\Big[\frac{\int^{\infty}_{R_{cn}}r^{2\lambda}\Phi^2_{nlj}(r)dr}
                {\int^{\infty}_0r^{2\lambda}\Phi^2_{nlj}(r)dr}\Big]^{1/\lambda}
\end{equation}
with $\lambda=1,2$, where $\Phi_{nlj}$ is the radial wavefunction of valence nucleon 
in the orbital ($nlj$). The $D_1$ represents the probability of the valence nucleon 
outside the range of the interaction radius $R_{cn}$, and $D_2$ gives the contribution 
of the outer part to the rms radius. Here we suggest to use the strict halo definition, 
i.e. $D_1 > 50$\%. Moreover, $D_2$ are usually larger than 90\% in the halo cases. 
It means that the uncertainty caused by the inner part is quite small. Very recently, 
Carstoiu et al.~\cite{cars01} calculated the density distribution of last proton in 
$^8$B and extracted its rms radius. They found that $D{_1(2)}=0.29(0.85)$ for $R_N=4$ 
fm whereas $D{_1(2)}=0.64(0.90)$ for $R_N=2.5$ fm. Taking the latter case, they judged 
that $^8$B is a proton halo nucleus. Obviously, $D_{1(2)}$ values vary with $R_N$. 
In order to avoid the arbitrariness, $R_N$ should be fixed to the radius of interaction 
potential, i.e. $R_N \equiv R_{cn}$. With the same procedure of Carstoiu et al., 
we gained $<r^2>^{1/2}=4.5$ fm and $D{_1(2)}=0.40(0.80)$ with $R_{cn}=3.6$ fm. Thus, 
$^8$B is the nucleus with thick skin in nature.

\quad With the experimental binding energies, angular momenta, spins, etc., 
the Sch{\"o}rdinger equations are numerically solved and the rms radii are calculated 
for a number of single-particle states by means of the single-particle potential 
model. In this model, the shapes of nuclear potentials are adopted Woods-Saxon form 
and the Coulomb and spin-orbit-coupling potentials are considered. The relations 
between the rms radii and separation energies are illustrated for different angular 
momenta. The new scaling laws and necessary conditions for neutron halos and proton 
halos are established, respectively. The actual conditions are more looser than those 
extracted by the square well, due to the effects of short-distance behaviours of the 
valence nucleons at the edges of interaction potentials. Because of the existence of 
Coulomb barrier, some resonant states possibly have proton halo structures. The region 
of true proton halos existence is pointed out. Moreover, the contributions of outer 
parts are estimated by the radii of interaction potentials. The halo state can be 
strictly judged by this estimation.

\quad This work was supported by the Major State Basic Research Development Program under 
Grant No. G200077400, and the National Natural Science Foundation of China under Grant 
Nos. 19875087, 10075077 and 10105016. One of us, C.J. Lin wish to thank to Prof. Z.Y. Ma 
and Prof. Z.X Li for many useful discussion.

\newpage
\begin{figure}
\vspace*{0.5cm}
\caption{\label{fig1}
The single-particle wavefunctions for two bound states $2s_{1/2}$ and $1p_{1/2}$ 
of $^{11}$Be with different shapes. The radius of interaction potential is also 
indicated.}
\end{figure}

\begin{figure}
\vspace*{0.5cm}
\caption{\label{fig2}
The rms radii of valence neutrons vary with the separation energies for a number of 
states in light neutron-rich nuclei. The solid circles, squares and triangles represent
the results of $l$ = 0, 1, 2 states, respectively. The solid lines are the fit results
with the origins of Woods-Saxon potentials and the dashed lines are the results 
extracted by square potentials.}
\end{figure}

\begin{figure}
\vspace*{0.5cm}
\caption{\label{fig3}
Same as the Fig. 2, but for the proton cases. The typical states of $^{17}$F(2s$_{1/2}$), 
$^{8}$B(1p$_{3/2}$) and $^{17}$F(1d$_{5/2}$) are respectively pointed out.}
\end{figure}

\begin{figure}
\vspace*{0.5cm}
\caption{\label{fig4}
The shadow area illustrates the region of true proton halos occurrence, including the 
bound states and resonance states. The location of $^{17}$F(2s$_{1/2}$) state is 
indicated.}
\end{figure}


\begin{thebibliography}{99}
\bibitem{tani01}
I. Tanihata, H. Hamagaki, O. Hashimoto, Y. Shida, N. Yoshikawa, K. Sugimoto, O. Yamakawa, 
T. Kobayashi, and N. Takahashi, Phys. Rev. Lett. 55 (1985) 2676.

\bibitem{tani02}
I. Tanihata, H. Hamagaki, O. Hashimoto, S. Nagamiya, Y. Shida, N. Yoshikawa, O. Yamakawa, 
K. Sugimoto, T. Kobayashi, D.E. Greiner, N. Takahashi, and Y. Nojiri, Phys. Lett. B 160 (1985)380.

\bibitem{fuku01} 
M. Fukuda, T. Ichihara, N. Inabe, T. Kubo, H. Kumagai, T. Nakagawa, Y. Yano, I. Tanihata, 
M. Adachi, K. Asahi, M. Kouguchi, M. Ishihara, H. Sagawa, and S. Shimoura, Phys. Lett. B 
268 (1991) 339.

\bibitem{bazi01}
D. Bazin, B.A. Brown, J. Brown, M. Fauerbach, M. Hellstrom, S.E. Hirzebruch, J.H. Kelley, 
R.A. Kryger, D.J. Morrissey, R. Pfaff, C.F. Powell, B.M. Sherrill, and M. Thoennessen,
Phys. Rev. Lett. 74 (1995) 3569.

\bibitem{tani03}
I. Tanihata, T. Kobayashi, T. Suzuki, K. Yoshida, S. Shimoura, K. Sugimoto, K. Matsuta, 
T. Minamisono, W. Christie, D. Olson, H. Wieman, Phys. Lett. B 287 (1992) 307.

\bibitem{suzu01}
T. Suzuki, R. Kanungo, O. Bochkarev, L. Chulkov, D. Cortina, M. Fukuda, H. Geissel, M. Hellstrom, 
M. Ivanov, R. Janik, K. Kimura, T. Kobayashi, A.A. Korsheninnikov, G. Munzenberg, F. Nickel, 
A.A. Ogloblin, A. Ozawa, M. Pfutzner, V. Pribora, H. Simon, B. Sitar, P. Strmen, K. Sumiyoshi, 
K. Summerer, I. Tanihata, M. Winkler, and K. Yoshida, Nucl. Phys. A 658 (1999) 313. 

\bibitem{labi01}
M. Labiche, N.A. Orr, F.M. Marques, J.C. Angelique, L. Axelsson, B. Benoit, U.C. Bergmann, 
M.J.G. Borge, W.N. Catford, S.P.G. Chappell, N.M. Clarke, G. Costa, N. Curtis, A. D'Arrigo, 
E.de Goes Brennand, O. Dorvaux, G. Fazio, M. Freer, B.R. Fulton, G. Giardina, S. Grevy, 
D. Guillemaud-Mueller, F. Hanappe, B. Heusch, K.L. Jones, B. Jonson, C. Le Brun, S. Leenhardt, 
M. Lewitowicz, M.J. Lopez, K. Markenroth, A.C. Mueller, T. Nilsson, A. Ninane, G. Nyman, 
F. de Oliveira, I. Piqueras, K. Riisager, M.G. Saint Laurent, F. Sarazin, S.M. Singer, O. Sorlin, 
and L. Stuttge, Phys. Rev. Lett. 86 (2001) 600.

\bibitem{zhuk01}
M.V. Zhukov, A.A. Korsheninnikov, and M.H.Smedberg, Phys. Rev. C50 (1994) R1.

\bibitem{smed01}
M.H. Smedberg, T. Baumann, T. Aumann, L. Axelsson, U. Bergmann, M.J.G. Borge, D. Cortina-Gil, 
L.M. Fraile, H. Geissel, L. Grigorenko, M. Hellstrom, M. Ivanov, N. Iwasa, R. Janik, B. Jonson, 
H. Lenske, K. Markenroth, G. Munzenberg, T. Nilsson, A. Richter, K. Riisager, C. Scheidenberger, 
G. Schrieder, W. Schwab, H. Simon, B. Sitar, P. Strmen, K. Summerer, M. Winkler, and M.V. Zhukov,
Phys. Lett. B 452 (1999) 1.

\bibitem{fuku02}
M. Fukuda, M. Mihara, T. Fukao, S. Fukuda, M. Ishihara, S. Ito, T. Kobayashi, K. Matsuta, 
T. Minamisono, S. Momota, T. Nakamura, Y. Nojiri, Y. Ogawa, T. Ohtsubo, T. Onishi, A. Ozawa, 
T. Suzuki, M. Tanigaki, I. Tanihata, and K. Yoshida, Nucl.Phys. A 656 (1999) 209.

\bibitem{guim01}
V. Guimaraes, J.J. Kolata, D. Peterson, P. Santi, R.H. White-Stevens, S.M. Vincent, F.D. Becchetti, 
M.Y. Lee, T.W. O'Donnell, D.A. Roberts, and J.A. Zimmerman, Phys. Rev. Lett. 84 (2000) 1862.

\bibitem{ozaw01}
A. Ozawa, T. Kobayashi, H. Sato, D. Hirata, I. Tanihata, O. Yamakawa, K. Omata, K. Sugimoto, 
D. Olson, W. Christie, and H. Wieman, Phys. Lett. B 334 (1994) 18.

\bibitem{suzu02} 
T. Suzuki, H. Geissel, O. Bochkarev, L. Chulkov, M. Golovkov, D. Hirata, H. Irnich, Z. Janas, 
H. Keller, T. Kobayashi, G. Kraus, G. Munzenberg, S. Neumaier, F. Nickel, A. Ozawa, 
A. Piechaczek, E. Roeckl, W. Schwab, K. Summerer, K. Yoshida, and I. Tanihata, 
Nucl. Phys. A 616 (1997) 286c. 

\bibitem{navi01}
A. Navin, D. Bazin, B.A. Brown, B. Davids, G. Gervais, T. Glasmacher, K. Govaert, P.G. Hansen, 
M. Hellstrom, R.W. Ibbotson, V. Maddalena, B. Pritychenko, H. Scheit, B.M. Sherrill, M. Steiner, 
J.A. Tostevin, and J. Yurkon, Phys. Rev. Lett. 81 (1998) 5089.

\bibitem{liu01}
Z.H. Liu, C.J. Lin, H.Q. Zhang, Z.C. Li, J.S. Zhang, Y.W. Wu, F. Yang, M. Ruan, 
J.C. Liu, S.Y. Li, and Z.H. Peng, Phys. Rev. C 64 (2001) 034312.

\bibitem{lcj01}
Lin Cheng-Jian, Liu Zu-Hua, Zhang Huan-Qiao, Wu Yue-Wei, Yang Feng, and Ruan Ming,
Chin. Phys. Lett. 18 (2001) 1183 and 1446.

\bibitem{hans01}
P.G. Hansen and B. Jonson, Europhys. Lett. 4 (1987) 409.

\bibitem{hans02}
P.G. Hansen and A.S. Jensen, Annu. Rev. Nucl. Part. Sci. 45 (1995) 591.

\bibitem{fedo01}
D.V. Fedorov, A.S. Jensen, and K. Riisager, Phys. Rev. C 49 (1994) 201; 
Phys. Rev. C 50 (1994) 2372; Phys. Lett. B 312 (1993) 1.

\bibitem{jens01}
A.S. Jensen, K. Riisager, Phys. Lett. B 480 (2000) 39.

\bibitem{sher01}
R. Sherr, Phys. Rev. C 54 (1996) 1177.

\bibitem{satc01} 
G.R. Satchler, {\it Direct Nuclear Reactions}, Oxford University Press, Oxford OX2 6DP (1983).

\bibitem{ozaw02}
A. Ozawa, T. Kobayashi, T. Suzuki, K. Yoshida, and I. Tanihata1, Phys. Rev. Lett. 84 (2000) 5493.

\bibitem{cars01}
F. Carstoiu, L. Trache, C.A. Gagliardi, R.E. Tribble, and A.M. Mukhamedzhanov, 
Phys. Rev. C 63 (2001) 054310.

\end{thebibliography}
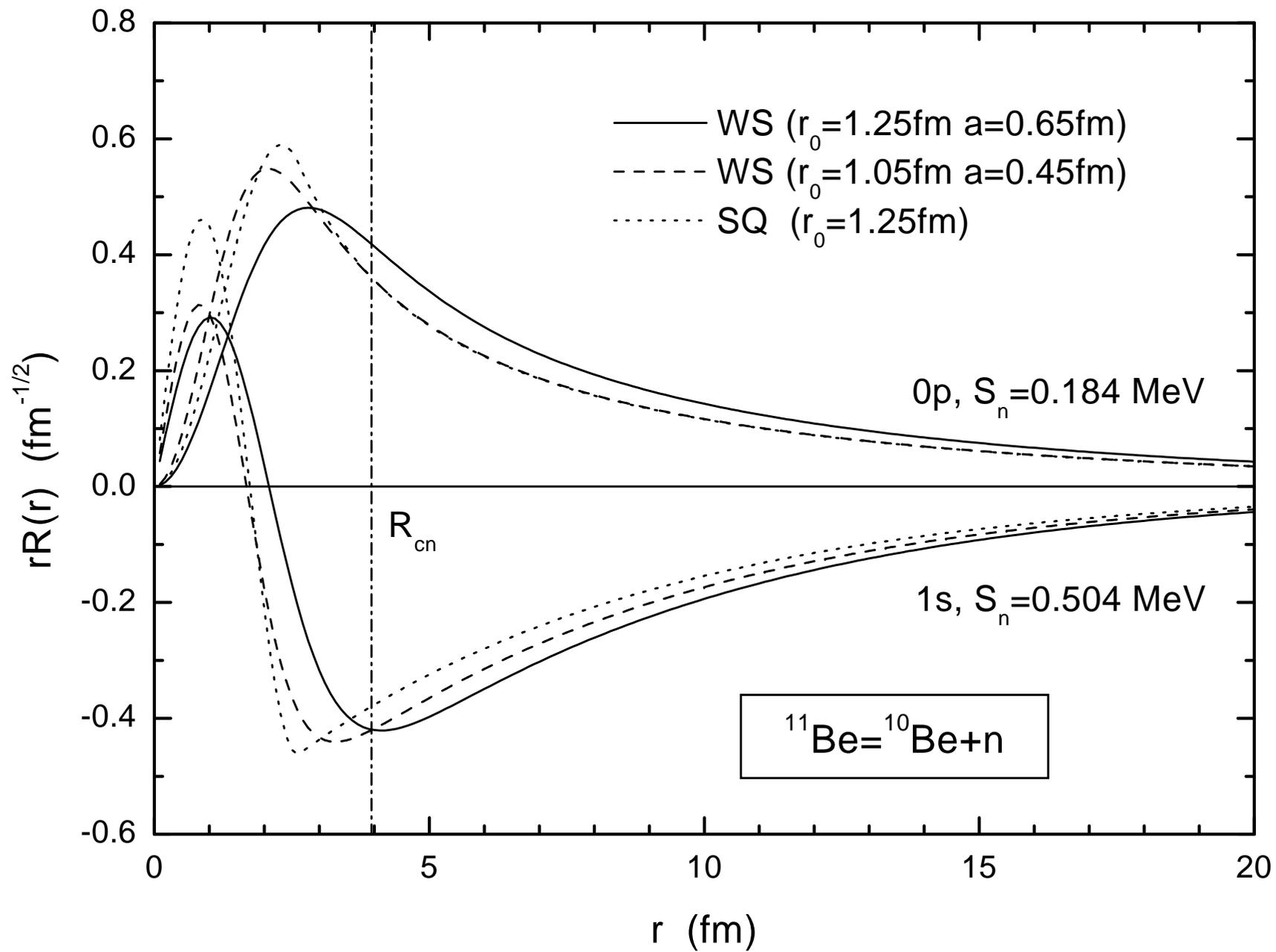
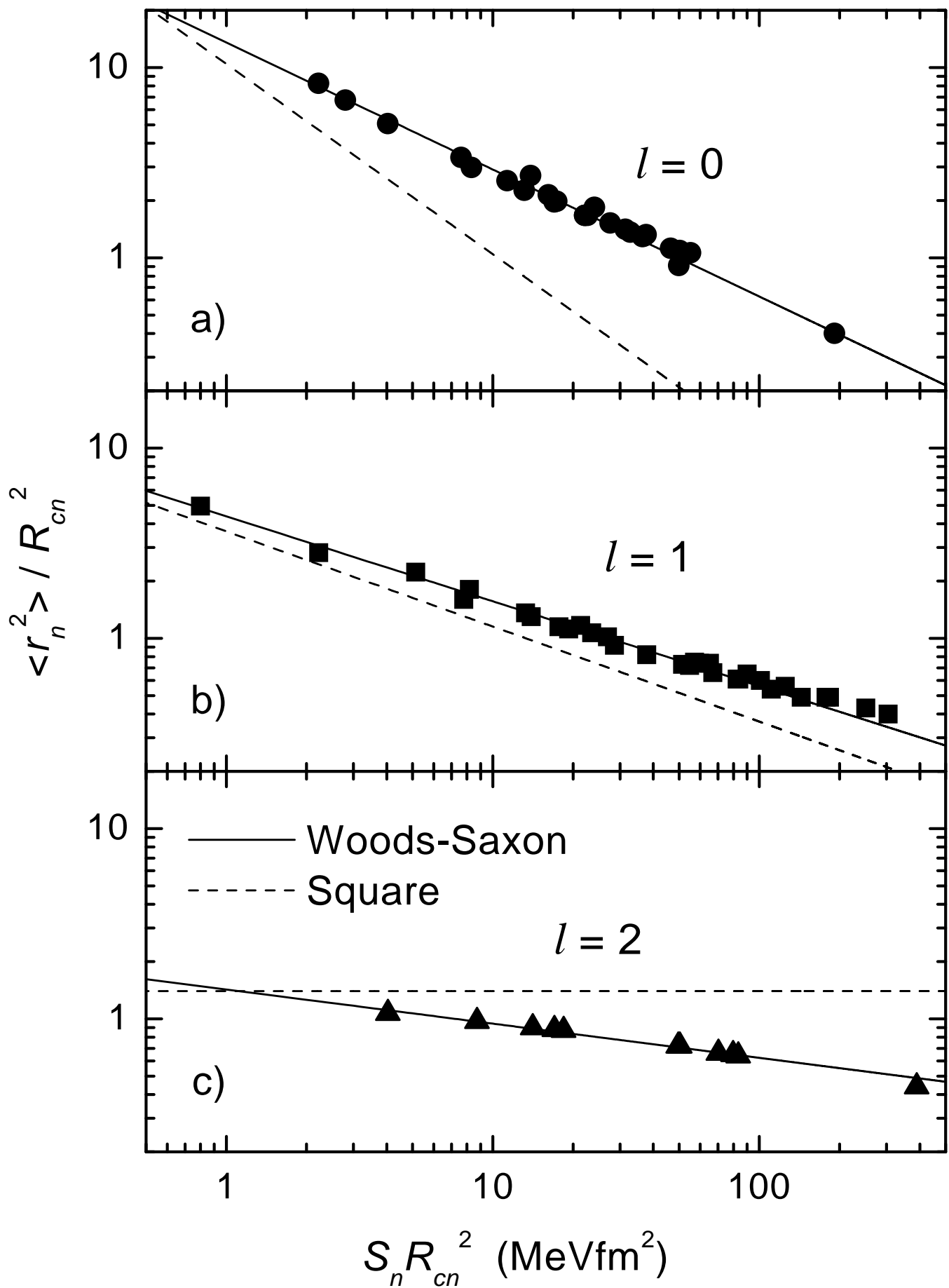
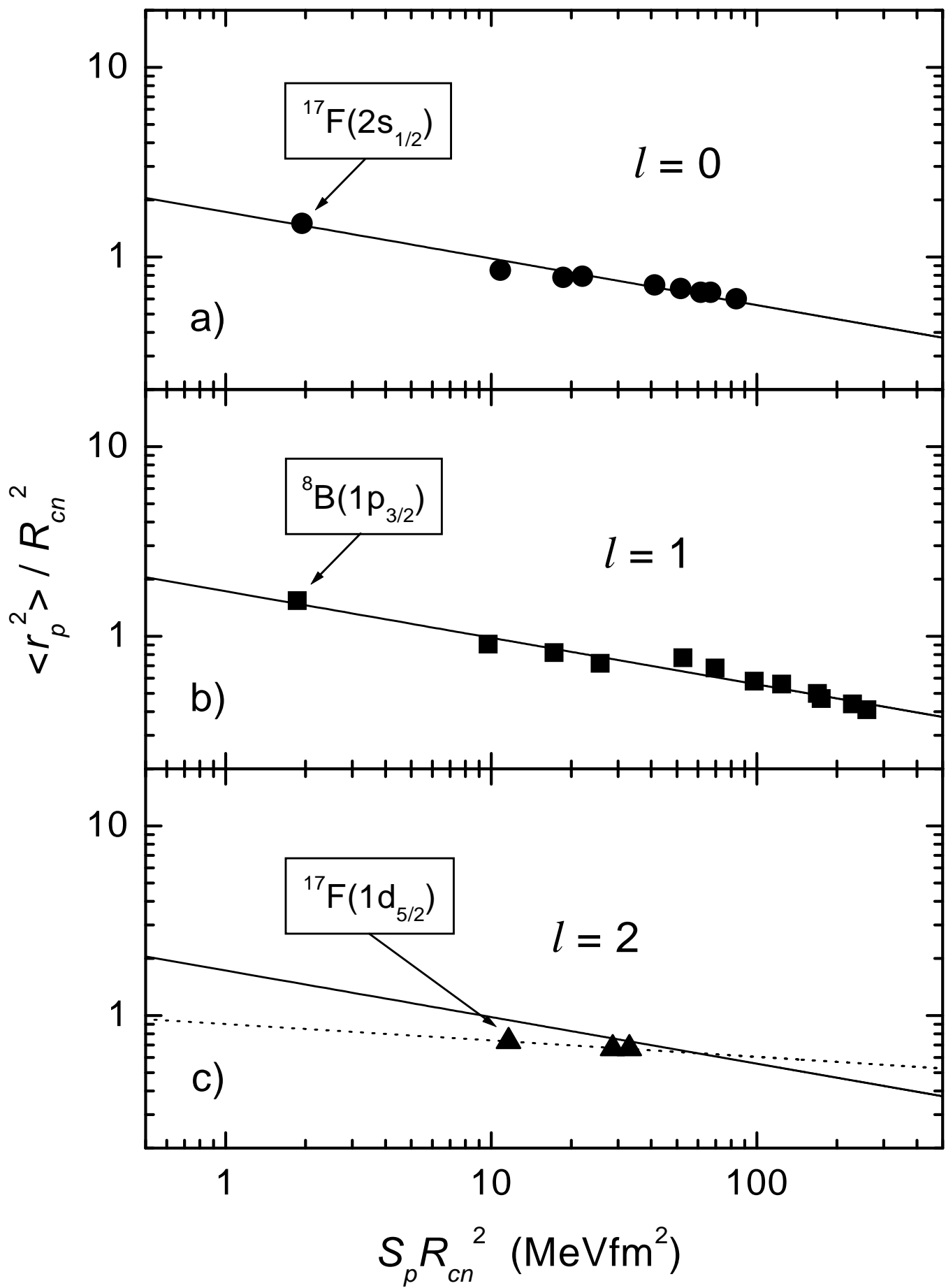
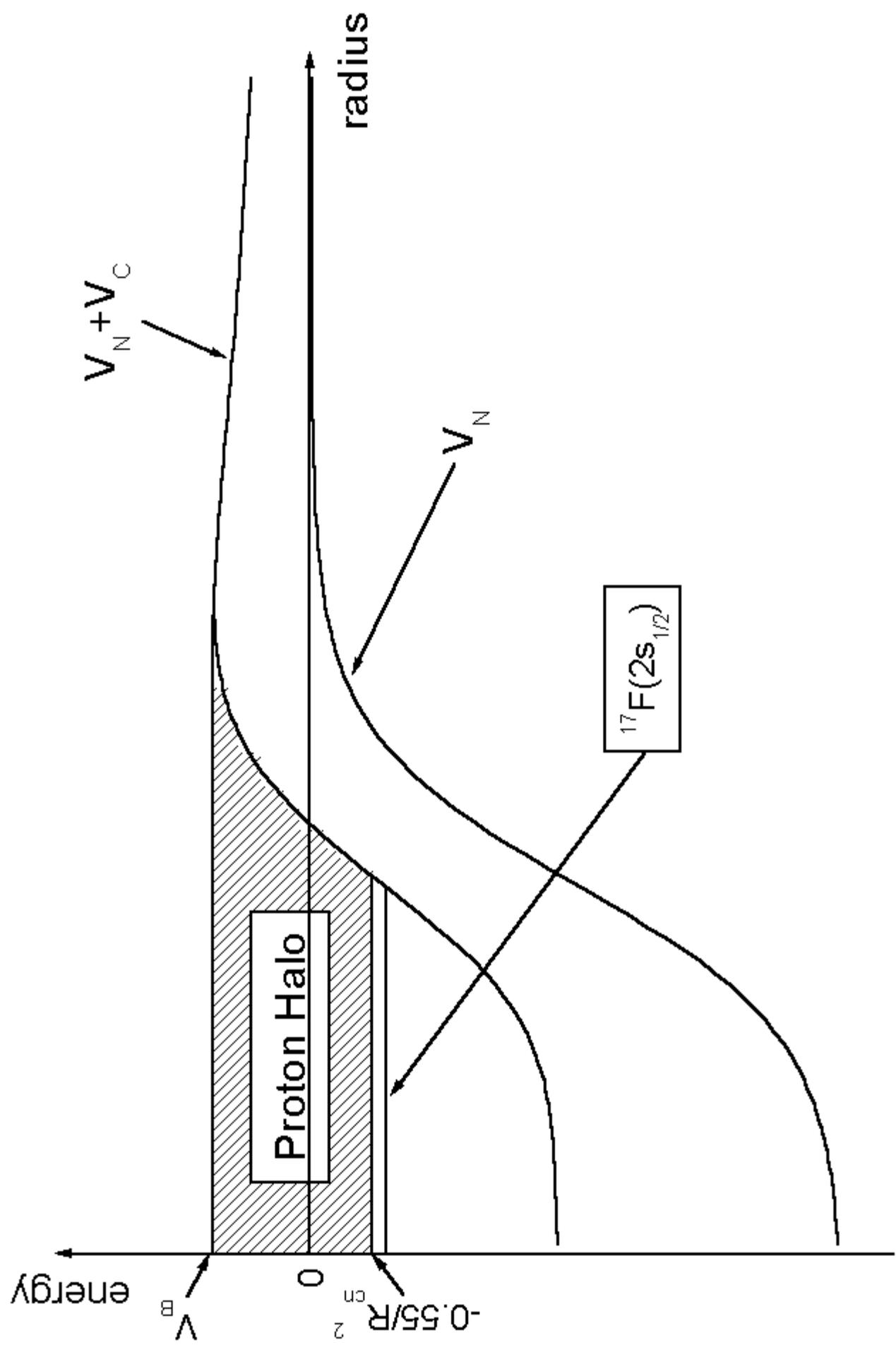
\end{document}